\documentclass[preprint]{revtex4}

\usepackage{dcolumn}
\usepackage{graphicx}
\usepackage{bm}
 
\begin{document} 
 
\title{Field-linked States of Ultracold Polar Molecules} 
\date{\today}
\author{A. V. Avdeenkov}
\author{D. C. E.  Bortolotti}
\author{J. L. Bohn}
\email{bohn@murphy.colorado.edu}
\affiliation{JILA and Department of Physics,
University of Colorado, Boulder, CO 80309-0440}

\begin{abstract} 
We explore the character of a novel set of ``field-linked'' states
that were predicted in [A. V. Avdeenkov and J. L. Bohn, Phys. Rev. Lett. {\bf 90},
043006 (2003)].  These states exist at ultralow temperatures in the presence 
of an electrostatic field, and their properties are strongly
dependent on the field's strength.  We clarify the nature of these 
quasi-bound states by
constructing their wave functions and determining their approximate quantum
numbers.  As the properties of field-linked states are strongly defined by 
anisotropic dipolar  and Stark interactions, we  construct adiabatic surfaces 
as functions of both the intermolecular distance and the angle that the
intermolecular axis makes with the electric field.  Within an adiabatic approximation 
we solve the 2-D Schr\"{o}dinger equation to find bound states, whose energies
correlate well with resonance features found in fully-converged
multichannel scattering calculations.
\end{abstract}  

 
\maketitle 
 
\section{Introduction} 
In the modern world of physics,  manipulation of 
quantum phenomena in atoms and molecules forms the basis for future 
applications.   With the development of new techniques for cooling 
and trapping polar molecules, new opportunities to harness them  
appeared \cite{Bethlemreview,Weinstein,Egorov,Bethlem1,Bethlem2,
Bochinski, Kerman,Rangwala,ShaferRay}.
In particular, the interactions between
pairs of molecules are likely to  be susceptible to manipulation in an
electric field.  This in turn may imply an ability to direct the
course of chemical reactions \cite{Bala}, to influence the many-body
physics of degenerate Bose or Fermi gases composed of polar molecules
\cite{shlyap,Lushnikov,Derevianko,baran,Goral}, or to manipulate 
quantum bits \cite{DeMille}.

A particularly attractive opportunity for controlling
intermolecular interactions emerges in a set of novel
long-range bound states of molecular pairs \cite{AB1,AB2}.
In the presence of an external electric field, the 
counterplay between Stark and dipole-dipole 
interactions generates shallow potentials that are predicted 
to support bound states of two polar molecules.   For OH molecules 
we have estimated that the bound states do not exist at all 
for fields below about 1000 V/cm \cite{AB1}.
Thus the field plays an essential role in binding the molecules into 
an [OH]$_2$ dimer; we have accordingly dubbed this new kind of molecular 
state a ``field-linked'' state. The purpose of this
communication is to further clarify the structure of field-linked (FL) states. 
Interestingly, quadrupolar interactions between
metastable alkaline-earth atoms exhibit similar states in the presence
of magnetic fields \cite{Derevianko2,Santra,Kokoouline}.

Schematically, the FL states originate in avoided crossings
between a pair of potential energy curves: one that represents an attractive
dipolar interaction converging to a high-energy Stark threshold; and
one that represents a repulsive dipolar interaction converging to
a lower-energy threshold.  The characteristic size of the FL states
is therefore roughly determined by equating the dipolar energy $\mu^2/R^3$
to the Stark energy $\mu {\cal E}$.  Here $R$ is the distance between the
molecules, $\mu$ is their dipole moment, and ${\cal E}$ is the
field strength.  The length scale of the avoided crossing is then
$R_{scale} = (\mu / {\cal E})^{(1/3)} \approx 1250 {\cal E}^{-1/3}$
for a ``typical'' dipole moment of 1 Debye, where 
$R_{scale}$ is measured in units of $a_0$ (the Bohr radius) and
${\cal E}$ is measured in V/cm.  Thus for a reasonable-sized laboratory field
of $10^4$ V/cm, the size of the FL state is $\sim 60 a_0$, although
extremely weakly bound states can be far larger than this.

Ref. \cite{AB1} described the FL states in this simple curve-crossing
picture.  Adiabatic potential curves for the OH-OH interaction were
constructed by expanding the relevant potential into partial waves in
the intermolecular coordinate.  For clarity, only the lowest partial 
waves, $L=0,2$ were included.  While intuitively appealing, this
picture is inadequate, and indeed a partial wave expansion is
inappropriate, for the following reason.
The dipole-dipole interaction can strongly couple different values of $L$, 
with a strength on the order of $\approx \mu^2/R^3$.  At the typical 
scale distance 
$R_{scale}$, the dipole coupling exceeds the centrifugal interaction by a 
ratio $2m\mu^2/\hbar^2 R_{scale}$, where $m$ is the reduced mass of the 
molecular pair.  For our example case of $\mu=1$ D, ${\cal E} = 10^4$ V/cm, 
and for a light molecule (like OH) with a reduced mass $m = 10$, this ratio 
is already $\approx 100$.  The ratio becomes even larger in a stronger field, 
or for a heavier molecule.  Therefore $L$ is no longer a good quantum number 
for the FL states, but rather the relative orientation of the molecules is of 
more significance.   

Accordingly, in this paper we present a formulation of FL states in
terms of  potential energy surfaces in $(R, \theta)$, where 
$\theta$ is the angle that the intermolecular axis makes
with respect to the electric field.  Within an adiabatic representation,
we compute FL states as bound states of a single surface.  Qualitatively,
these identify the FL states as confined to a narrow range
about $\theta = 0$, so that their motion consists primarily
of vibration along the field axis.  Additionally, we show that the
binding energies predicted by this adiabatic
approximation agree remarkably well with resonance positions
determined from fully-converged multichannel scattering
calculations.

\section{Model} 

Because the FL states are generated primarily by the competition between
Stark and dipolar interactions, our model will focus almost exclusively
on these two terms in the Hamiltonian.  In particular, our simplifying 
assumptions here are:

1) The individual molecules are assumed to be in their electronic ground
states, to be rigid rotors, and to lie in their rotational ground states.
It is assumed that none of these degrees of freedom can be excited at 
the large intermolecular separations and low relative energies that we 
consider.

2) Each molecule is assumed to have total spin $j$ and to have a non-$\Sigma$
electronic ground state that can support a lambda-doublet.  Again, at the 
intermolecular separations, energies, and fields of interest, it is assumed that
$j$ is approximately conserved.  We ignore hyperfine structure in the model,
so that $j$ is an integer for bosonic molecules, and a half-integer for
fermionic molecules.  While hyperfine structure 
is well-known to be important in ultracold collisions, it is not germane to 
the main discussion of dipolar interactions, and can in any event be included 
in a straightforward way later. 

3) The projection of each molecule's angular momentum onto its own interatomic
axis, denoted $\omega$, takes only the two values $\pm j$.  As a point
of comparison, the energy difference between the $j=3/2$, $|\omega| = 3/2$ 
ground state and the $j=3/2$, $|\omega| = 1/2$ excited state of OH is 270 K
\cite{Coxon}, so this restriction is not such a bad one.  

4) We work in the limit of large electric field, i.e., in the
linear Stark regime where the electric field interaction dominates 
the lambda-doublet splitting.
Thus the molecular states are characterized by
the signed quantities $\pm \omega$, rather than linear combinations of
$\omega$ and $-\omega$ characteristic of the zero-field limit.
We will describe some effects of lambda-doubling in the following,
but they will be perturbative in this limit.
A readable account of molecular wave functions in this approximation 
is given in Ref. \cite{Schreel}.

5) Finally, we assume that the molecules never get close enough
together for short-range interactions such as hydrogen bonding,
exchange, or chemical reactions, to contribute.  In addition,
we neglect long range interactions such as dispersion and quadrupole-quadrupole
interactions, as being negligible compared to dipole-dipole interactions.

Although this model does not describe any particular molecule, it lays
the groundwork for constructing FL states for any desired molecule.
To keep the magnitudes of observable quantities realistic in the following, 
we use as model parameters the dipole moment (1.68 D), lambda-
doublet splitting (0.055 cm$^{-1}$), and mass (17 amu) of the OH radical.

\subsection{Basis set}

Within the simplifications outlined above, the internal state of
an individual rigid-rotor molecule is specified by three quantum numbers: 
$j$, $\omega$, and the projection of the molecule's angular momentum on
an appropriate external axis.  to describe the Stark interaction
this axis is conveniently taken as the electric field axis.  However,
to describe FL states we choose instead to quantize this angular momentum
along the intermolecular axis.
This emphasizes the dimer nature of the FL states
and allows a reasonable description of how the dipole-dipole 
forces act ultimately to keep the molecules from crashing
into one another.

Each molecule ($i=1,2$) is thus described by a rigid rotor  wave function, 
\begin{equation} 
\langle \bm{\hat{e}}_i|j k_i \omega_i \rangle = \sqrt{ {2j+1 \over 8\pi^2}} 
D^{j*}_{k_i \omega_i}(\alpha_i,\beta_i,\gamma_i), 
\end{equation} 
where notation for the electronic wave function is suppressed, under the 
assumption that it plays no role at the temperatures and electric fields of 
interest.  Here $k_i$ and $\omega_i$ are the projections of total 
angular momentum $j$ onto the intermolecular axis and onto the molecule's 
own body-frame axis, respectively.  The Euler angles 
$\bm{\hat{e}}_i=(\alpha_i,\beta_i,\gamma_i)$ are 
referred to the intermolecular axis. 
We further couple the molecular spins into a total spin $J$:
\begin{eqnarray} 
&&\langle \bm{\hat{e}}_{1},\bm{\hat{e}}_{2}|(1,2)JK \rangle   = \nonumber \\
&& \;\;\;\; \sum_{k_{1}k_{2}} 
\langle \bm{\hat{e}}_{1}|j k_1 \omega_1 \rangle 
\langle \bm{\hat{e}}_{2}|j k_2 \omega_2 \rangle
\langle j_1 k_1 j_2 k_2 | J K \rangle.
\end{eqnarray} 
Here we introduce the shorthand notation $(1,2)$ do denote the
internal molecular quantum numbers $(j_1\omega_1,j_2\omega_2)$.

As for the relative motion of the molecules, we wish to avoid an expansion
into partial waves, as mentioned in the Introduction.  We thus
consider a basis set for the complete wave function
\begin{eqnarray}
\Psi^{\cal M}_{(1,2)JK}(R,\theta,\phi,\bm{\hat{e}}_1,\bm{\hat{e}}_2) &=&
{1 \over \sqrt{2\pi}} \exp (i{\cal M}\phi) F^{\cal M}_{(1,2)JK}(R,\theta)
\nonumber \\
&\times& \langle \bm{\hat{e}}_{1},\bm{\hat{e}}_{2}|(1,2)JK \rangle,
\end{eqnarray}
where the $F$'s are as-yet-unspecified functions of $(R,\theta)$.
The projection of the total angular momentum onto the electric field
axis, ${\cal M}$, is the only rigorously conserved quantity in the
Hamiltonian for FL states; we therefore separate it at the outset.
It will affect the functions $F$ via centrifugal energies.

In addition, the wave functions must incorporate the proper symmetry
under the exchange of identical molecules, denoted by the operator
${\hat P}_{12}$.  The symmetrized states are constructed in Appendix A,
and define a pair of quantum numbers $s$ and $x$:
\begin{eqnarray}
\label{xsym}
{\hat P}_{12} F^{{\cal M},s}_{(1,2)JK} &&= s F^{{\cal M},s}_{(1,2)JK},
\nonumber \\
{\hat P}_{12} |(1,2)JK \rangle_x &&= x |(1,2)JK \rangle_x .
\end{eqnarray}
The quantities $s$ and $x$ are not separately conserved by the Hamiltonian,
but must satisfy the constraint
\begin{equation}
sx = \Biggl\{ \begin{array}{l}
+1 \;\;\;\; {\rm for} \;\; {\rm  bosons} \\
-1 \;\;\;\; {\rm for} \;\; {\rm fermions}. \\
\end{array} 
\end{equation}

Finally, it is useful to consider the effect of the parity operator 
${\hat I}$ that inverts all coordinates through the system's center-of-mass.  
Eigenvalues $\epsilon$ of this operator are obviously not conserved 
by the electric field, yet we can
construct basis sets that are eigenfunctions of ${\hat I}$, as is done 
in Appendix A.  When we consider matrix elements of the electric field
and dipole-dipole Hamiltonia, we find that the quantity
\begin{equation}
\label{qdef}
q \equiv \epsilon s (-1)^K
\end{equation}
is rigorously conserved (see Appendix B).  Our completely general basis
then takes the form
\begin{equation}
\label{basis}
\Psi^{{\cal M},q}_{(s,x)(1,2)JK} = 
{1 \over \sqrt{2\pi}} \exp(i{\cal M}\phi) F^{{\cal M},s}_{(1,2)JK} 
|(1,2)JK \rangle_{x,q},
\end{equation}
whose  explicit representation in terms of unsymmetrized 
basis functions is given in Appendix A.

\subsection{Hamiltonian matrix elements} 

To uncover the joint motion in $(R,\theta)$ that governs the FL states,
we will expand the total wave function into the basis (\ref{basis}) and
integrate over all other  degrees of freedom to derive a set of 
coupled-channel differential equations for the functions $F$.  In this
section we therefore construct the Hamiltonian matrix elements in the
``internal'' basis $|(1,2)JK \rangle_{x,q}$.

Ignoring the exchange, quadrupole-quadrupole and dispersion interactions
 as we did in \cite{AB1}, our model Hamiltonian can be written as 
\begin{eqnarray} 
H = \sum_{i=1,2} \left( T_i + H^{S}_i \right)
+V_{\mu \mu}, 
\end {eqnarray} 
where $T_i$ and  $H^{S}_{i}$ are the translational kinetic energy 
Stark energy of each molecule, 
and $V_{\mu \mu}$ is the dipole-dipole interaction. 

In the following subsections we list the matrix elements of the various
terms of the Hamiltonian in the unsymmetrized basis.  Transformation
into the symmetrized basis set is accomplished in Appendix B.

\subsubsection{Stark Interaction} 

An electric field with strength ${\cal E}$ that points along the positive
$z$-axis in the laboratory frame will have spherical components ${\cal E}_q$
in the reference frame that rotates with the intermolecular axis.
 The relation between the two is given by a Wigner rotation matrix:
\begin{equation}
\label{Stark_rotated}
{\cal E}_{q} = {\cal E}D^1_{0q}(\phi,\theta,0).
\end{equation}
The components of the molecular dipole moment $\bm{\mu}$  
can be written in terms of reduced spherical harmonics $C_{1q}(\alpha,\beta)$
where, as above, $\alpha $ and $\beta$ are Euler angles relative to the
intermolecular axis.
The Stark Hamiltonian for a single molecule is then 
\begin{equation} 
\label{Stark_Ham} 
-\bm{\mu} \cdot \bm{{\cal E}} = -\mu {\cal E} 
\sum_q (-1)^q C_{1q}(\alpha,\beta)D^1_{0-q}(\phi,\theta,0) 
\end{equation}

The integration over each molecule's internal coordinates 
yields, for the unsymmetrized basis set (and remembering that 
$j_{1}=j_{2}=j$ )\cite{BS},
\begin{eqnarray} 
\label{matrixS} 
&& \langle (1,2)JK| H^S |(1',2')J'K'\rangle= \nonumber \\
&& \;\;\;\; -\mu {\cal E} (-1)^{j+K'} [j]^{2}[J][J'] 
\delta_{1,1'}\delta_{2,2'}
 d^{1}_{0,K'-K}(\theta) 
\nonumber   \\ 
&& \;\;\;\; \times \left\{ 
 \begin{array}{ccc} 
  J' & 1 & J \\ 
  j & j & j \\
 \end{array} \right\} 
 \left( 
 \begin{array}{ccc} 
  J & 1 & J' \\ 
  K & K'-K & -K' 
 \end{array} \right) 
 \nonumber 
 \\  
&& \;\;\;\; \times \Biggl[ (-1)^{\omega_1} 
 \left( 
 \begin{array}{ccc} 
  j & 1 & j \\ 
 -\omega_1 & 0 & \omega_1 
 \end{array} \right) 
\nonumber \\
&& \;\;\;\;\;\;\;\; +(-1)^{\omega_2+J+J'} 
 \left( 
 \begin{array}{ccc} 
  j & 1 & j \\ 
 -\omega_2 & 0 & \omega_2 
 \end{array} \right) 
 \Biggr] ,
 \\ 
\end{eqnarray}
where $[y] \equiv \sqrt{2y+1}$.

\subsubsection{Dipolar Interaction} 
 
The dipole-dipole interaction reduces to a particularly simple form in the 
rotating frame: 
\begin{eqnarray} 
\label{Dipole_Ham} 
V_{\mu \mu} &=& { \bm{\mu}_1 \cdot \bm{\mu}_2 
- 3(\bm{{\hat R}}\cdot \bm{\mu}_1)(\bm{{\hat R}}\cdot \bm{\mu}_2) 
\over R^3}    \nonumber \\ 
&=& - {\sqrt{6} \over R^3} \left[ \mu_1 \otimes \mu_2 \right]^2_0. 
\end{eqnarray} 
Here $\left[ \mu_1 \otimes \mu_2 \right]^2_0$ is the (2,0) component of the second-rank 
tensor formed by the product of $\bm{\mu}_1$ and $\bm{\mu}_2$.  The zero refers to 
the cylindrically symmetric component around the intermolecular axis. 
 
Following a treatment similar to the Stark effect above, we note that 
\begin{equation} 
\left[ \mu_1 \otimes \mu_2 \right]^2_0 = \mu^2 
\sum_q C_{1q}(\alpha_1,\beta_1) C_{1-q}(\alpha_2,\beta_2) 
\langle 1q 1-q | 20 \rangle. 
\end{equation}  
Now the angular integration over each molecule's internal coordinates is 
again straightforward, yielding 
\begin{eqnarray} 
\label{dipoleunsym}
&&\langle (1,2)JK | V_{\mu \mu} | (1',2') J'K' \rangle = \nonumber \\
&& \;\;\;\; -\frac{\mu^2}{R^3}\sqrt{30}(-1)^{K'-\omega_1 -\omega_2}[J][J'][j]^4
\nonumber 
\\ 
&& \;\;\;\; \times \left( 
 \begin{array}{ccc} 
  j & 1 & j \\ 
 -\omega_1 & 0 & \omega_1 
 \end{array} \right) 
 \left( 
 \begin{array}{ccc} 
  j & 1 & j \\ 
 -\omega_2 & 0 & \omega_2 
 \end{array} \right) 
\nonumber \\
&& \;\;\;\; \times \left( 
 \begin{array}{ccc} 
  J & 2 & J' \\ 
  -K & 0 & K' 
 \end{array} \right) 
  \left\{ 
 \begin{array}{ccc} 
 J'& 2 & J \\ 
 j & 1 & j \\ 
 j & 1 & j
 \end{array} \right\}.
\end{eqnarray} 
This matrix element is independent of the orientation $\theta$, as
it must be.
 
\subsubsection{Kinetic energy} 
 
The centrifugal Hamiltonian in the rotating frame is no longer 
diagonal, but rather couples states with $K,K \pm 1$ projections. 
Within our basis it is more 
convenient to present the angular momentum operator as
\cite{launay} 
\begin{eqnarray} 
\hat{l}^{2}=\hat{\cal J}^{2}+\hat{J}^{2}-2\hat{J}_z-(\hat{J}_{-}\hat{\cal J}_{+}+\hat{J}_{+}\hat{\cal J}_{-}) 
\end{eqnarray} 
Knowing that 
\begin{eqnarray} 
\hat{J}^{2}|J(1,2)K>&=&J(J+1)|J(1,2)K> ,
\nonumber \\ 
\text{and}
\hat{J}_{\pm}|J(1,2)K>&=& \sqrt{J(J+1)-K(K \pm 1)} \nonumber \\
&& \times | J(1,2)K \pm 1> 
\end{eqnarray} 
and using the definition of $\hat{F}^{2}$ and $\hat{F}_{\pm}$ 
\cite{varsh} we have 
\begin{eqnarray} 
\hat{l}^{2}\exp(i{\cal M} \phi)F^{\cal M}_{(1,2)JK}(R,\theta) |J(1,2)K>= 
\nonumber 
\\ 
\times e^{i{\cal M} \phi} \Biggl(|J(1,2)K> 
\hat{A}_0(K) 
+|J(1,2)K-1>\hat{A}_{-1}(K) 
\nonumber \\ 
+|J(1,2)K+1>\hat{A}_{+1}(K) \Biggr) F^{\cal M}_{(1,2)JK}(R,\theta), 
\nonumber
\end{eqnarray} 
where 
\begin{eqnarray} 
\hat{A}_0(K)&=&-\frac{\partial^2}{\partial 
\theta^{2}}-\cot(\theta)\frac{\partial}{\partial \theta}+ 
\frac{1}{\sin^2(\theta)}{\cal M}^2 \nonumber \\
&& \;\;\;\; +J(J+1)-2K^2 \nonumber 
\\ 
\hat{A}_{\pm 1}(K)&=&-\frac{1}{\sqrt{2}}\sqrt{J(J+1)-K(K \mp 1)} 
\nonumber \\
&& \;\;\;\; \times \biggl(-\frac{\partial}{\partial \theta} \pm 
{\cal M} \frac{1}{\sin(\theta)}\biggr) 
\end{eqnarray} 
For convenience, we will in the following neglect the Coriolis-type
couplings ${\hat A}_{\pm 1}$.  Like many other perturbations,
these can be incorporated later, if necessary.

\subsection{Schr\"{o}dinger equation}
Within our scheme we have the following Schr\"{o}dinger equation 
\begin{eqnarray} 
\label{schrod} 
&-& \Biggl( \frac{\hbar^2}{2m}\frac{\partial^2}{\partial R^{2}}+E 
\Biggr)f_{i}(R,\theta) \nonumber \\
&+&\sum_{i'}(\hat{V}^{cent}_{i,i'}(R,\theta) + V_{i,i'}(R,\theta))
f_{i'}(R,\theta)=0 
\end{eqnarray} 
where  $i=\{{\cal M}, (1,2)JK,x,q \}$ 
and $f_{i}=F_{i}/R$ .
\begin{widetext} 
Solutions of the coupled-channel partial differential equations
(\ref{schrod}), subject to scattering boundary conditions, yield
both the energies and resonance widths of the FL states.  To
clarify the nature of these states, however, we first invoke
a Born-Oppenheimer approximation.  Thus we will diagonalize 
the model Hamiltonian for fixed values of the pair $(R,\theta)$,
and seek bound state in one of the resulting potentials.
In a single adiabatic surface $V^{adiab}(R,\theta)$, 
the Schr\"{o}dinger equation reads
\begin{eqnarray} 
\label{schrod2} 
\Biggl( - \frac{\hbar^2}{2m} \Biggl( \frac{\partial^2}{\partial R^{2}}+ 
\frac{1}{R^2}\biggl( \frac{\partial^2}{\partial 
\theta^{2}}+\cot(\theta)\frac{\partial}{\partial \theta}- 
\frac{1}{\sin^2(\theta)}M_F^2 -J(J+1)+2K^2 \biggr)  \Biggr) 
\nonumber 
\\ 
+V^{adiab}(R,\theta)-E \Biggr)f(R,\theta)=0 
\end{eqnarray} 
\end{widetext}

\section{Characteristics of the field-linked states} 

For concreteness, we consider here a pair of bosonic molecules
with $j=1$, and parameters corresponding to the OH radical, as 
discussed above.  From the similar model in Ref. \cite{AB1},
we then expect to see a small number of FL states at modest
electric field values.  Our aim in this section
is to describe these states approximately in terms of the
quantum numbers in our basis set defined in the previous section.

\subsection{Adiabatic surfaces} 

The number of adiabatic potential surfaces is set by the
number of internal states of the molecules.  (Contrast this
to an expansion in partial waves, where the number of channels
is, in principle, infinite.)  For
a pair of $j=1$ molecules, the present model contains 36 channels,
hence 36 surfaces.  Moreover, conservation of $q$ implies that these 36
surfaces split into two sets of 18 channels each.  The surfaces
for $q=1$ and $q=-1$ are  identical, if only the Stark and
dipolar interactions are included, as we assume.  We find that
including  the lambda-doublet interaction leaves the $q=-1$ surfaces
unchanged, but introduces some weak avoided crossings among the
$q=1$ surfaces.  
Since lambda-doubling is a perturbation for the fields we consider, 
we will ignore this small effect.  Hereafter we report on the
$q=-1$ surfaces.  Additionally, the quantities $\omega_i$ are
conserved in the absence of lambda-doubling, meaning that we can 
further classify the surfaces according to whether $\omega_1=\omega_2$
or $\omega_1 = -\omega_2$. 

\begin{figure}
\includegraphics[angle=-90,width=3in]{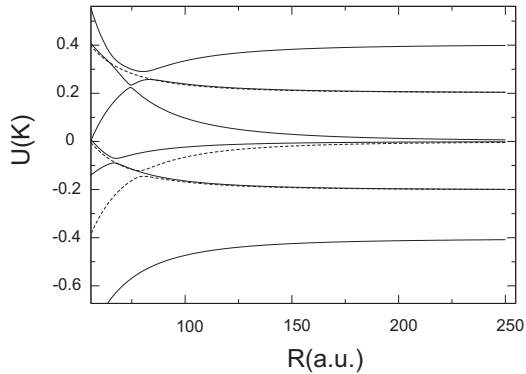}
\caption{
A ``slice'' through the adiabatic potential energy surfaces for
an electric field strength ${\cal E}=10^4$ V/cm.  In these surfaces
$q=-1$, $\omega_1 = \omega_2$, and even (solid line) and odd
(dashed line) values of $\langle J \rangle$ are distinguished.}
\label{slice}
\end{figure}

Subdividing the  surfaces in this way yields nine surfaces with 
$q=-1$ and $\omega_1=\omega_2$, which are of greatest interest
here.  Slices through these surfaces at a fixed angle $\theta = 5^{\circ}$
are shown in Figure \ref{slice}.  Here we take the applied electric 
field strength to be ${\cal E}=10^4$ V/cm.
Empirically, we find that surfaces with
even values of $\langle J \rangle$ (solid lines) are only
weakly coupled to surfaces with odd values of $\langle J \rangle$
(dashed lines).  This consideration further reduces the number
of surfaces necessary to describe the FL states.

\begin{figure}
\includegraphics[width=3in]{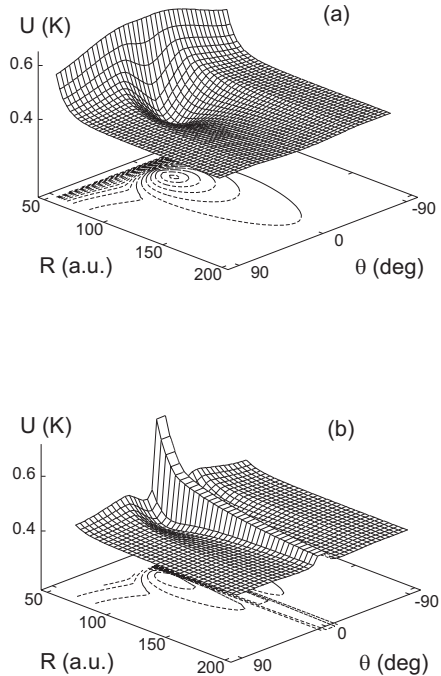}
\caption{
Adiabatic potential energy surfaces corresponding to the highest-lying
slice in Fig. 1.  Shown are the ${\cal M}=0$ (a) and ${\cal M}=1$
(b) cases.}
\label{mf0}
\end{figure}

The FL states are bound states of the highest-lying surface in
Figure \ref{slice}, which is clearly generated by avoided crossings.
The complete surface in the $(R,\theta)$ plane is shown in Figure \ref{mf0}
for both the rotationless case ${\cal M}=0$ and a rotating
case with ${\cal M}=1$.  Addition of the centrifugal energy makes the 
${\cal M}=1$ surface substantially more shallow than the ${\cal M}=0$ 
surface; in fact we find six bound states for ${\cal M}=0$, 
and only two for ${\cal M}=1$ (see Table I).

\begin{table} 
\begin{tabular}{|c|c|d|}  \hline
 ${\cal M}$ & $v$  &  \text{Energy (K)}  \\   \hline
 0 & 0 &  0.0282    \\ \hline
 0 & 1 &  0.00550    \\ \hline
 0 & 2 &  0.000455    \\ \hline
 1 & 0 &  0.00545     \\ \hline
\end{tabular}  
\caption{ Binding energies in Kelvin of FL states.  Each state
is identified by its rotation ${\cal M}$ about the electric field
axis, and by a vibrational quantum number $v$.  These energies
refer to states even under the reflection $\theta \rightarrow
\pi - \theta$.  Additional states, odd under this symmetry,
are separated in energy by less than several $\mu$K from the
ones listed.}
\end{table} 

To gain a better understanding of the nature of the FL states,
it is useful to evaluate mean values of the quantum numbers in our basis set.
In general, the symmetry-type quantum numbers $x$, $s$, and $\epsilon$
are badly nonconserved, and average to zero.  
However, the angular momentum quantum numbers $J$ and  $K$
typically have well-defined mean values that are useful
for interpretation.

\begin{figure}
\includegraphics[angle=-90,width=3in]{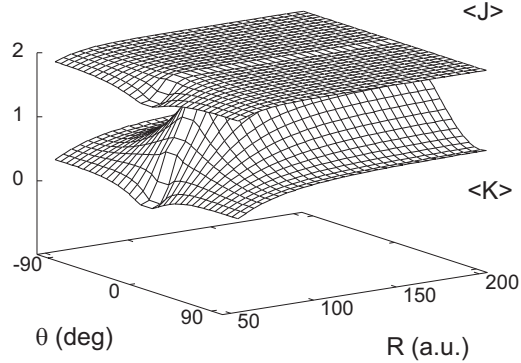}
\caption{
Average values of $J$ and $K$  for the ${\cal M}=0$ potential in
Fig.2. Note that the orientation of the axes is different from
that in Figs 2 and 4.}
\label{JK}
\end{figure}

Figure \ref{JK} shows surface plots of the mean values $\langle J \rangle$
and $\langle K \rangle$ for the FL potential surface (note that the 
axes are rotated relative to Fig. \ref{mf0}).  Near the minima
of the potential wells, we find that $\langle J \rangle \approx 
\langle K \rangle \approx 2$.
These values characterize the FL states at large separation $R$.
Ref. \cite{Schreel} presents a simple and useful semiclassical
picture of the dipole's orientation in the OH molecule. In this model
the dipole precesses around the molecule's total angular momentum
$\bm{j}$, and on average points along $\bm{j}$ when $\omega>0$,
and against $\bm{j}$ when $\omega<0$.  Thus when 
$\langle J \rangle \approx \langle K \rangle \approx 2$ and
$\omega_1 \ \omega_2$, as is the case here, the
dipole moments are both aligned on average in the same direction, 
roughly along the intermolecular axis, and hence attract one another.

At smaller values of $R$, $\langle J \rangle$ remains nearly equal to
2, but $\langle K \rangle$ drops all the way to 0.  This reflects the
influence of the avoided crossings in the surfaces.  
Again invoking a semicalssical picture, $\langle J \rangle = 2$,
$\langle K \rangle=0$
implies that the dipole moments are now aligned roughly 
perpendicular to the intermolecular axis, in a side-by-side 
orientation where they repel one
another.  This is the reason the FL state is stable against collapse to 
smaller $R$. 

The avoided crossings that allow FL states to be supported
have their origin in the fact the the Stark 
interaction is diagonal in the laboratory frame (defined by the
field axis), whereas the dipolar interaction is diagonal in the
rotating  frame (defined by the intermolecular axis).  Competition between
these two symmetries generate the avoided crossings.  However, in the
limit where $\theta \rightarrow 0$ the two axes coincide and both
interactions become diagonal in $K$.  In this case the avoided crossings
become diabatic crossings, and there is a conical intersection in 
the surfaces.  Our description in terms of adiabatic surfaces is, therefore,
incomplete.  It is however useful, as we will see in the next section.
There may be interesting information on geometrical
phases inherent in the FL states; this will be a topic of future study.

\subsection{Bound states} 

\begin{figure}
\includegraphics*[width=6in]{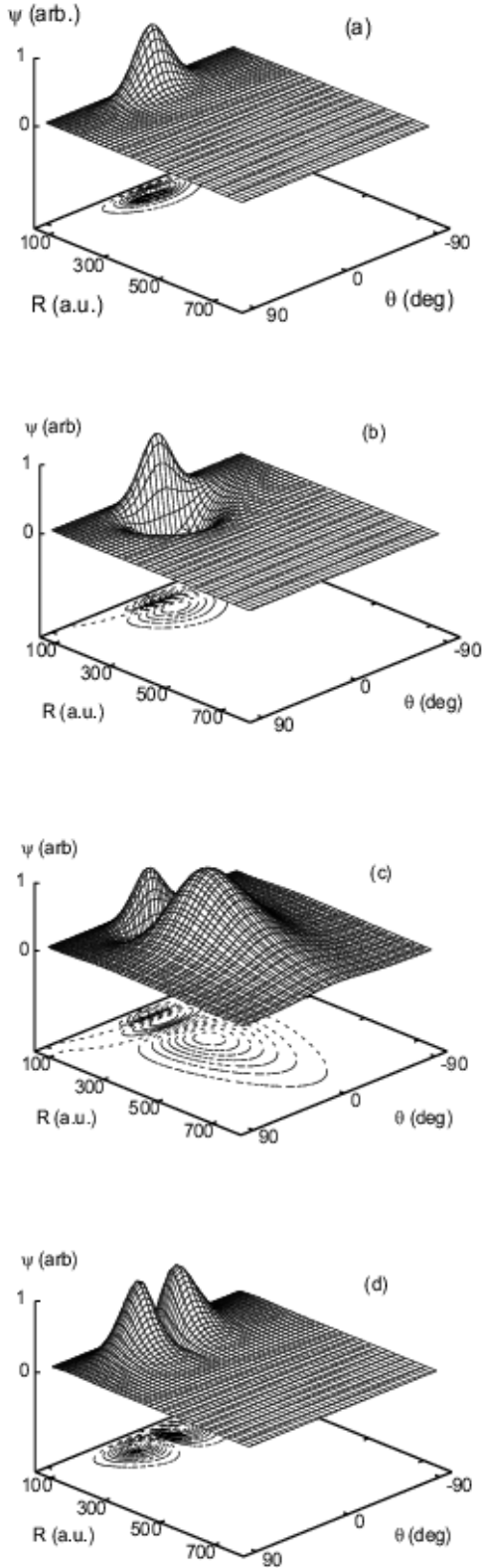}
\caption{Wave functions of FL states, for the potential surfaces
shown in Fig. 2.  For ${\cal M}=0$, there are three vibrational
states, $v=0$ (a), $v=1$ (b), and $v=2$ (c).  For ${\cal M}=1$,
there is a single state, with $v=0$ (d). }
\label{wfs}
\end{figure}

To complete a description of the FL states we must understand their motion
in $R$ and $\theta$.  Each bound state is nearly doubly-degenerate with respect
to reflection in the $\theta=\pi/2$ plane.  In Figure \ref{wfs} we present
wave function plots of those bound states  that have
even reflection symmetry, corresponding to the bound states listed
in Table I.  In this figure, (a-c) refer to the
${\cal M}=0$ case, and (d) to the ${\cal M}=1$ case.
for ${\cal M}=0$, it is immediately evident that these states
exhibit zero-point motion in the $\theta$ direction, and that excitations
are primarily in the $R$ direction.  We therefore label the FL states with
a vibrational quantum number $v$.  For ${\cal M}=1$, a nodal line
appears along the $\theta=0$ direction, owing to the centrifugal
energy that forces the molecules away from the electric field axis.

In realistic laboratory circumstances, the FL states are quasi-stable,
being subject to dissociation into free molecules in lower-energy internal
states \cite{AB1}.  Nevertheless, the adiabatic bound states we have
identified here correspond to real features of these dissociating states.
To show this, we have carried out a complete coupled-channel scattering 
calculation in a laboratory-frame representation, similar to that in
Ref. \cite{AB2}, but without including hyperfine structure.  We have 
included partial waves up to $L=16$ to ensure convergence at the several
percent level in scattering observables.  

\begin{figure}
\includegraphics*[width=3.5in]{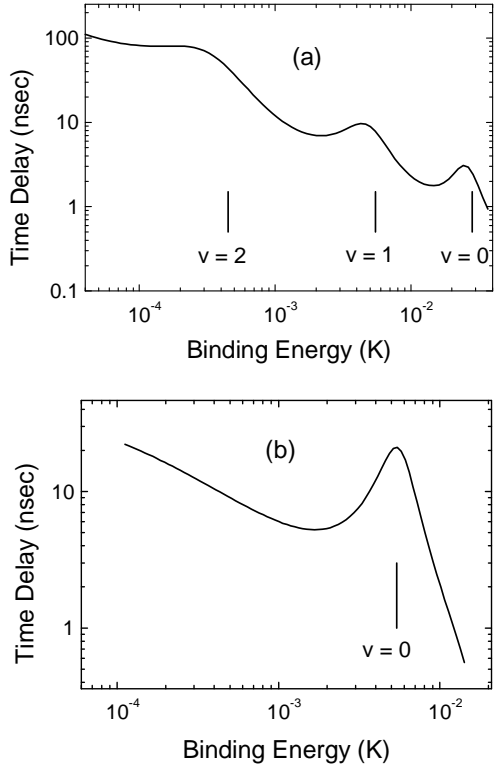}
\caption{ Time delay as defined in Eqn. (22) for the ${\cal M}=0$ (a)
and ${\cal M}=1$ (b) FL states.  Resonance peaks appear at characteristic
energies that correlate well with the binding energies as determined
from the FL adiabatic surfaces (vertical lines).}
\label{delay}
\end{figure}

We compute the time delay for the scattering process, defined as
\cite{Fano}
\begin{equation}
\tau = 2 \hbar {d \delta \over dE},
\end{equation}
where $\delta$ is the eigenphase sum, i.e., the sum of the inverse
tangents of the eigenvalues of the scattering K-matrix.  This quantity, plotted
in Figure \ref{delay}, exhibits  peaks at energies where resonances
occur.  There is also a significant background component, arising
from threshold effects, but the
peaks are nevertheless visible.  Also indicated are the binding energies 
of the FL states as given in Table I.  The good agreement 
between the 
two calculations suggests that when FL resonances are observed in experiments,
the nature of the resonant states will be well-approximated by the
wave functions determined above.

In general, resonances in ultracold polar molecule scattering will
come in three varieties.  The ``true'' field-linked states, like the ones that
we describe here, are largely independent of physics at small
values of $R$.  We can verify this assertion by changing the small-$R$
boundary conditions in our multichannel scattering calculation.  The
positions of the FL resonances do not depend at all on these boundary 
conditions.  However, their lifetimes can fluctuate within a factor of 
$\sim 2$, since the continuum states into which they can decay
do depend on short-range physics.

A second type of FL state appears to have components at both large and
small $R$.  Examples of these are found for states lying below
the middle threshold in Fig. \ref{slice}.  We find that their positions are
relatively insensitive to the short-range boundary conditions, but that
their lifetimes vary wildly.  We refer to these as ``quasi-FL'' states.
Finally a third category of resonance is strongly sensitive to 
initial conditions, both in position and width.  These are resonant states
of the short-range interaction, which are expected to be numerous
in low-energy molecular collisions \cite{Forrey,AB3}.

\section{Outlook}

We have left out many details of molecular structure and 
interactions, in order to emphasize the basic structure
of the field-linked states.  This structure is remarkably
simple, and consists primarily of a pair of molecules 
in relative vibrational motion along an axis that nearly
coincides with the direction of the electric field.  The number
of FL states is not large, since the forces holding them
together are necessarily weak.  

Significantly, to adapt this simple picture to a particular
molecular species requires only a detailed knowledge of the 
structure of each molecule
separately, plus some information on long-range parameters
such as dispersion coefficients.  In other words, realistic modeling of
experimentally probed FL states can probably be achieved
using currently existing information.  This is in stark contrast
with molecular collisions involving close contact between the
molecules, in which case existing  potential energy
surfaces are likely to be inadequate for to describe collisions
at ultralow temperatures.
 
\begin{acknowledgements}
This work was supported by the NSF and an ONR-MURI grant, 
and by a grant from the W. M. Keck
Foundation.  We acknowledge illuminating discussions with J. Hutson.
\end{acknowledgements}

\appendix

\section{Symmetrized wave functions}

To incorporate the effects of symmetrization under the exchange 
(${\hat P}_{12}$) and parity (${\hat I}$) operations, we follow 
the treatment of Alexander and DePristo 
\cite{alexander}.  To this end it is convenient to relate the
Euler angles of each molecule to the electric field axis rather than the
intermolecular axis; these Euler angles are denoted $\bm{\hat{e}}^L$.
The symmetry operations then perform the following functions:
\begin{eqnarray}
{\hat P}_{12} : && \bm{R} \rightarrow -\bm{R} \nonumber \\
&& \bm{\hat{e}}^L_1 \rightarrow \bm{\hat{e}}^L_2 \nonumber \\
&& \bm{\hat{e}}^L_2 \rightarrow \bm{\hat{e}}^L_1 . \nonumber
\end{eqnarray}
\begin{eqnarray}
{\hat I} : && \bm{R} \rightarrow -\bm{R} \nonumber \\
&& \bm{\hat{e}}^L_1 \rightarrow {\hat I}(\bm{\hat{e}}^L_1) \nonumber \\
&& \bm{\hat{e}}^L_2 \rightarrow {\hat I}(\bm{\hat{e}}^L_2) . \nonumber
\end{eqnarray}
The last two lines imply that ${\hat I}$ acts on each
molecule by inverting the molecule's coordinates through its
own center of mass.

The effect of particle exchange on the internal coordinates is
determined by making the explicit rotation to the lab frame:
\begin{eqnarray}
&& {\hat P}_{12} \langle \bm{\hat{e}}_1, \bm{\hat{e}}_2 | (1,2)JK \rangle
\nonumber \\
&& = {\hat P}_{12} \sum_{m_{12}} 
\langle \bm{\hat{e}}^L_1,\bm{\hat{e}}^L_2 | (1,2)Jm_{12} \rangle
D^J_{m_{12},K} (\phi,\theta,0)
\nonumber \\
&& = \sum_{m_{12}} 
(-1)^{2j+J}
\langle \bm{\hat{e}}^L_1,\bm{\hat{e}}^L_2 | (2,1)Jm_{12} \rangle
(-1)^J D^J_{m_{12},-K} (\phi,\theta,0)
\nonumber \\
&&= (-1)^{2j} \langle \bm{\hat{e}}_1, \bm{\hat{e}}_2 | (2,1)J-K \rangle.
\nonumber
\end{eqnarray}
Here we have used the reflection symmetry of the Wigner $D$ functions,
\begin{eqnarray}
D^J_{mK}(\pi + \phi, \pi - \theta, 0) = 
(-1)^J D^J_{m-K}(\phi, \theta, 0)  \nonumber
\end{eqnarray}
and the usual exchange symmetry of the Clebsch-Gordan coefficients.
Similarly the relative wave functions transform as
\begin{eqnarray}
&&{\hat P}_{12} \left[ \exp(i{\cal M}\phi) 
F^{\cal M}_{(1,2)JK}(R,\theta) \right] \nonumber \\
&& \;\;\;\; = 
(-1)^{\cal M}\exp(i{\cal M}\phi) F^{\cal M}_{(1,2)JK}(R,\pi-\theta).
\nonumber
\end{eqnarray}
An appropriately symmetrized basis for exchange is therefore
given by Eqn. (\ref{xsym}), where
\begin{eqnarray}
&&F^{{\cal M},s}_{(1,2)JK} = {1 \over 2} \nonumber \\
&& \;\;\;\; \times
\left[ F^{\cal M}_{(1,2)JK}(R,\theta) + s(-1)^{\cal M}
F^{\cal M}_{(1,2)JK}(R,\pi-\theta) \right] ,
\nonumber \\
&& |(1,2)JK \rangle_x = {1 \over \sqrt{2(1+\delta_{12}\delta_{K0})} }
\nonumber \\
&& \;\;\;\; 
 \times \left[ |(1,2)JK \rangle + x(-1)^{2j} |(2,1)J-K \rangle \right],
\nonumber
\end{eqnarray}
with $sx = \pm 1$ for bosons/fermions.

These basis functions can in turn be assembled into parity
eigenfunctions.  Note that ${\hat I}$ has the same effect on
the relative coordinates as does ${\hat P}_{12}$, so that 
$e^{i{\cal M}\phi} F^{{\cal M},s}_{(1,2)JK}$ is already a parity eigenstate
with eigenvalue $s$.  Denoting the parity of the total wave function
by $\epsilon$, the parity of the relative wavefunctions should be
$p = \epsilon s$, or $p =q (-1)^K$ in terms of our quantum number $q$ 
defined in Eqn.(\ref{qdef}).  This definition seems (and is)
completely arbitrary; it is justified by explicitly working out the
matrix elements for the Stark and dipole-dipole interactions, and
finding that both conserve the value of $q$.

The influence of ${\hat I}$ on each molecule is to reverse its direction 
of rotation about its own axis, and to introduce a phase \cite{Singer}:
\begin{eqnarray}
{\hat I} \langle \bm{\hat{e}}^L | j,m,\omega \rangle
= (-1)^{j-s} \langle \bm{\hat{e}}^L | jm -\omega \rangle. \nonumber
\end{eqnarray}
Because the phase factor is the same for each molecule, the action of
${\hat I}$ on the molecule pair is, by arguments similar to those
above,
\begin{eqnarray}
{\hat I} \langle \bm{\hat{e}}_1 \bm{\hat{e}}_2 | (1,2) JK \rangle
= (-1)^J \langle \bm{\hat{e}}_1 \bm{\hat{e}}_2 | (-1,-2) J-K \rangle,
\nonumber
\end{eqnarray}
where the notation $(1,2) \rightarrow (-1,-2)$ implies
$(j_1,\omega_1,j_2,\omega_2) \rightarrow $ 
$(j_1,-\omega_1,j_2,-\omega_2)$.  The symmetrized internal
basis function is then
\begin{eqnarray}
&&|(1,2)JK \rangle_{x,q} = \nonumber \\
&& \;\;\;\; {1 \over \sqrt{2} } \left[
|(1,2)JK \rangle_{x} + q(-1)^{J+K} |(-1,-2)J-K \rangle_{x}
\right]. \nonumber
\end{eqnarray}

\section{Conservation of $q$}

It is straightforward (if somewhat tedious) to write the
symmetrized matrix elements for different contributions to the
Hamiltonian, in terms of the unsymmetrized basis.  We present here some
of the key results, which rely mostly on the symmetry properties of
the angular momentum recoupling coefficients, as described in
Brink and Satchler \cite{BS}.

{\it Dipolar interaction.}  
In Eqn. (\ref{dipoleunsym}), the 9-$j$ symbol must be invariant
under exchanging its second and third rows, yet this operation 
introduces a phase shift $(-1)^{J+2+J'}$.  Therefore, we must have
$J+J'=$ even, and the matrix element (\ref{dipoleunsym}) is
invariant under the substitution
$(1,2) \rightarrow (2,1)$, $K \rightarrow -K$.  In the symmetrized
basis it reads
\begin{eqnarray}
&& _x \langle (1,2)JK | V_{\mu \mu} | (1',2')J'K' \rangle _x = 
\nonumber \\
 && \;\;\;\; \left( {1+xx' \over 2} \right) 
{1 \over \sqrt{ (1 + \delta_{12}\delta_{K0})(1 + \delta_{1'2'}\delta_{K'0}) }}
\nonumber \\
&& \;\;\;\; \times \biggl[ \langle (1,2)JK | V_{\mu \mu} | (1',2')J'K' \rangle 
\nonumber \\
&& \;\;\;\;\;\; + x'(-1)^{2j} \langle (1,2)JK | V_{\mu \mu} |  
(2',1')J'-K' \rangle \biggr].  \nonumber
\end{eqnarray}
Thus the exchange quantum number $x$ is explicitly conserved.
Similarly, the matrix elements are invariant under simultaneously 
reversing the signs of all $\omega$'s and $K$, whereby
\begin{eqnarray}
&& _{x,q} \langle (1,2)JK | V_{\mu \mu} | (1',2')J'K' \rangle _{x',q'} =
\nonumber \\
&& \;\;\;\; \biggl( {1 + qq'(-1)^{K+K'} \over 2} \biggr)
\nonumber \\
&& \;\;\;\; \times \biggl[ _{x} \langle (1,2)JK | V_{\mu \mu} 
| (1',2')J'K' \rangle _{x'} \nonumber \\
 &&\;\;\;\;\;\;\;\; +q'(-1)^{K'} \; _{x'} \langle (1,2)JK | V_{\mu \mu} 
| (-1',-2')J'-K' \rangle_{x'} \biggr]. \nonumber
\end{eqnarray}
Because $K=K'$ for the dipolar interaction, this implies in turn that
$q$ is conserved.  The matrix derivation of symmetrized matrix
elements for the lambda-doubling is exactly the same, and this
interaction also conserves $q$.

{\it Stark interaction.}
Symmetrized matrix elements of the Stark Hamiltonian (\ref{matrixS})
are slightly more complicated, since reversing the sign of $K$ also
affects the Wigner $d$-function.  Exploiting symmetries of the $d$ functions
yields
\begin{eqnarray}
&& _{x} \langle (1,2)JK | H^S | (1',2')J'K' \rangle_{x'} = \nonumber \\
&& \;\;\;\; \biggl( {1 - xx'(-1)^{K+K'} \over 2} \biggr)
{1 \over \sqrt{ (1 + \delta_{12}\delta_{K0})(1 + \delta_{1'2'}\delta_{K'0}) }}
\nonumber \\ 
&& \times \Biggl[ \langle (1,2)JK | H^S | (1',2')J'K' \rangle 
d^1_{0,K'-K}(\theta) \nonumber \\
&& \;\;\;\; +x'(-1)^{2j}
 \langle (1,2)JK | H^S | (2',1')J'-K' \rangle d^1_{0,K'-K}(\theta)
\Biggr]. \nonumber
\end{eqnarray}
In general, neither $x$, nor $K$, nor the product $x(-1)^K$, is conserved
by this part of the Hamiltonian.  However, the matrix elements in the
basis (\ref{basis}) become
\begin{eqnarray}
&& _{x,q} \langle (1,2)JK | H^S | (1',2')J'K' \rangle _{x',q'} =
\biggl( {1+ qq' \over 2} \biggr)
\nonumber \\ 
&& \times
\biggl[ _{x} \langle (1,2)JK | H^S | (1',2')J'K' \rangle _{x'}
\nonumber \\
&& \;\;\;\;\; + q'(-1)^{K'} \;
_{x'} \langle (1,2)JK | H^S | (-1',-2')J'-K' \rangle_{x'} \biggr], 
\nonumber
\end{eqnarray}
illustrating the conservation of $q$.

\begin{widetext}
{\it Centrifugal energy}.
Symmetrized over $x$, the centrifugal energy reads
\begin{eqnarray}
\label{cf}
_{x}<(1,2)JK|V^{cent}|(1',2')J'K'>_{x'}=
\frac{\hbar^2}{2mR^{2}}\delta_{J,J '}
\times \Biggl( {1 + xx' \over 2} \Biggr) \Biggl( {1 \over
\sqrt{4(1+\delta_{1,2}\delta_{K,0})(1+\delta_{1',2'}\delta_{K',0})} }\Biggr)
\nonumber
\\ \times
\Biggl[\biggl(\hat{A}_0(K)\delta_{K,K'}+\hat{A}_{-1}(K)\delta_{K,K'+1} + \hat{A}
_{+1}(K)\delta_{K,K'-1}\biggr)
\delta_{1,1'}\delta_{2,2'}
\\
\nonumber
\\
+x'\biggl(\hat{A}_0(K)\delta_{-K,K'}+\hat{A}_{-1}(K)\delta_{-K,K'+1} + \hat{A}_{
+1}(K)\delta_{-K,K'-1}\biggr)
\delta_{1,2'}\delta_{2,1'} \Biggr]
\nonumber
\\
\nonumber
\end{eqnarray}
From this point, translation into the $x,q$ symmetrized basis is
trivial.  In general, $q$ is not conserved by the Coriolis terms that
change $K$, but in the present treatment these terms are ignored.
\end{widetext}

\end{document}